# Smart Solution for the Detection of Preeclampsia

Iuliana Marin[1], Nicolae Goga[1,2], Andrei Vasilateanu[1], Alexandru Gradinaru[3], Vlad Racovita[4]

*Affiliation 1:* Faculty of Engineering in Foreign Languages, University Politehnica of Bucharest, Bucharest, Romania,
marin.iulliana25@gmail.com, andraevs@gmail.com
*Affiliation 2:* Molecular Dynamics Group, University of Groningen, Groningen, the Netherlands,
n.goga@rug.nl
*Affiliation 3:* Computer Graphics Department, University Politehnica of Bucharest, Bucharest, Romania,
alex.gradinaru@cs.pub.ro
*Affiliation 4:* SC Info World SRL, Bucharest, Romania, vlad.racovita@infoworld.ro

*Abstract*—This paper is written in the context of the international Eurostars project, i-bracelet. The main objective of the i-bracelet project - "Intelligent bracelet for blood pressure monitoring and detection of preeclampsia" is the creation of a portable medical device for uninterrupted monitoring of blood pressure and to detect the blood pressure problems (such as hypertension) and, in particular, preeclampsia. In the current paper is described the software component of this system used for monitoring, viewing and analyzing the blood pressure values coming from a smart bracelet developed in the context of the project. The software solution is available for Android and iOS phone users, as well as it is accessible from a browser. As a conclusion, the blood pressure of the future mothers should be monitored for living a safer and healthier life.

## I. INTRODUCTION

In this last 20 years, the medical domain undergone a big number of transformations due to the advances in technology. Lately, people are more interested in their health condition and spend more time to search for health related subjects and information. Technology influenced this greatly and helps to create value at the micro level of the medical domain [1].

This paper is related to the i-bracelet international Eurostars project. The main objective of the i-bracelet project - "Intelligent bracelet for blood pressure monitoring and detection of preeclampsia" is the development of a portable medical device for continuous monitoring of blood pressure and to detect the blood pressure problems (such as hypertension) and, in particular, preeclampsia.

Preeclampsia is a disorder which appears after 20 weeks of pregnancy. The main characteristics of preeclampsia are proteinuria and high blood pressure. The mean arterial pressure in the middle of the trimester demonstrated to be a good predictor of preeclampsia.

The main outcome of the project is an intelligent sensor system for the continuous management and ambulatory blood pressure monitoring. The sensor comprises a dynamic microfluidic layer which is located between two sensitive membranes. The microfluidic layer allows modifications of the impedance due to the deformation of the membrane that is subject to an external applied pressure. The high sensitivity of the sensor allows to obtain a complete signal of the waveform when the user's artery flattens due to the force applied on it as a result of the activities which are performed by the person.

The components of the bracelet are a sensor module, an electronic module and a wireless transmission module. The software application manages the system and provides data transmission and processing. The pressure sensor which detects the blood pressure waveform is based on a microfluidic carrier sensor which is placed on the wrist area. The measurement of the blood pressure is based on the arterial tonometry method. The electronic module provides conditioning and processing of the sensor signal, and the data is transmitted wirelessly to a smart phone or tablet. A dedicated algorithm of the created bracelet processes data and provides relevant physiological parameters that are displayed in numerical and graphical format on the display of the device.

Blood pressure is determined by analyzing the wrist artery activity with the help of the bracelet's sensors. The systolic and diastolic values are determined by analyzing the wavelength notches that are generated by sensors. An alarm signal is released when the blood pressure exceeds the normal level. Data is transferred to a computer for storage and further processing. The bracelet is useful for the pregnant women who can suffer from preeclampsia, namely high blood pressure and proteinuria. Until now there is no such product available on the market.

The process of developing an intelligent bracelet for arterial hypertension comprises: 1) early detection of preeclampsia, a common disorder among pregnant women, and 2) other diseases caused by hypertension like eclampsia, gestational hypertension and chronic hypertension. In the current paper is presented the software component of the i-bracelet system.

This paper is organized as follows. In the next section is presented the relevant literature review. In Section 3 is described the software component of the system, while in section 4 are drawn the conclusions.

## II. MATERIALS AND METHODS

High blood pressure is a frequent medical problem which appears during pregnancy, having a rate of occurrence which varies between 5 and 10% [2]. Hypertensive disorders that appear during pregnancy are classified as chronic hypertension, preeclampsia and eclampsia, preeclampsia overlapping on chronic hypertension and gestational hypertension which is a transient hypertension of pregnancy which appears during the second half of pregnancy [3].

Hypertensive diseases of pregnancy occurred for roughly 20 million women around the world in 2015 [4]. In the United States alone, pregnancy hypertensive diseases affect about 7% to 15% of the cases [5]. This resulted in 29,000 deaths in 2013. Blood pressure is determined by the blood force which acts upon the arteries walls.

Normally, blood pressure fluctuates throughout the day, ranging from 90 up to 119 mm Hg for the systolic value and from 60 up to 79 mm Hg for the diastolic value [6]. The blood pressure classification is determined by the values expressed in mm Hg for the systolic and diastolic values [7].

Almost 75 million American adults, namely 32% of the population, have hypertension, meaning 1 in every 3 adults [8]. About half of the analyzed persons have their health condition monitored. Just 2 out of 5 adults know that they suffer from hypertension.

Mild preeclampsia is hypertensive blood pressure that is greater than 140/90 mmHg on two occasions at a distance of at least 6 hours but without any evidence of organ lesions for a woman with a normal blood pressure before 20 weeks of gestation [9]. The syndrome is characterized by poor placental blood flow and a general process of the disease that can affect several organ systems [10].

The severe problems of preeclampsia include acute renal failure, cerebral edema, cerebral haemorrhage, seizures (edema), pulmonary edema, thrombocytopenia, haemolytic anemia, coagulopathy and hepatic injury [11]. The mean arterial pressure in the middle of the second quarter of the pregnancy period was found to be the best predictor of preeclampsia.

Current management plans for the prevention of eclampsia consist in the early disclosure of preeclampsia and the usage of therapy [12]. The prevention means are monitoring that can be done at the hospital or remotely, usage of therapy to maintain the blood pressure under a threshold, delivery to be done at the right time, prophylactic usage of magnesium sulphate during labor and prompt postpartum for the patients who have preeclampsia [13].

Most clinics and hospitals use aneroid or automated devices for measuring the blood pressure. Many pregnant women and certified personnel prefer the use of home blood pressure monitoring devices. However, the pregnancy data gathered is insufficient to guide a choice. Patients normally require education about the monitoring procedures and to interpret the blood pressure values returned by a certified medical device. Non-severely elevated blood pressure values have to be determined by repeated measurements which are taken at 15 minutes apart [14].

Blood pressure needs to be taken three times. The first value is neglected and the average of the second and third measurement defines the true blood pressure [15]. Many women with chronic hypertension give birth to healthy babies without big problems. Elevated blood pressure is a great risk for the mother and the infant [16]. Women with pre-existing chronic hypertension are at risk of developing serious complications such as gestational hypertension or preeclampsia which can cause mortalities if left untreated or undiagnosed [17].

Preeclamptic women exhibit certain symptoms such as swollen hands, feet and face, shortness of breath caused by the fluid in lungs, nausea or vomiting, abdominal pain, change in vision related to the loss of vision, blurred vision. Light sensitivity, excess of protein in urine (proteinuria) or other signs attributed to kidney failure are also some of the preeclampsia symptoms [18].

As diagnostic criteria, for the women who developed hypertension before pregnancy, the systolic values increase with over 30 mm Hg. The diastolic values are also affected, being with 15 mm Hg greater than usually [19]. The critical data peaks which exceed the average values are not be used.

The blood pressure is monitored during periods of time which last at least 6 hours. The presence of hypertension is declared if the blood pressure remains elevated. Proteinuria is detected when having over 0.3 grams of protein in a 24 hour urine sample [20]. The blood pressure measuring device which is available on the market for outpatient or home use is based on an inflatable cuff, making it inappropriate for prolonged wearing or for measuring during night time.

The current technologies used in existent smart bracelets that measure blood pressure are based on infrared and oximeter sensors which correlate the pulse rate and the blood oxygen level with blood pressure [21-23]. However, accuracy is a problem and such smart bracelets cannot be used in hospital settings. Other studies show a connection between the fetal heart rate and the uterine pressure [24]. The i-bracelet project solution which is illustrated in Fig. 1, comprises a system which gathers, interprets and displays information about the evolution of the blood pressure and heart rate based on the measurements performed by the created smart bracelet.

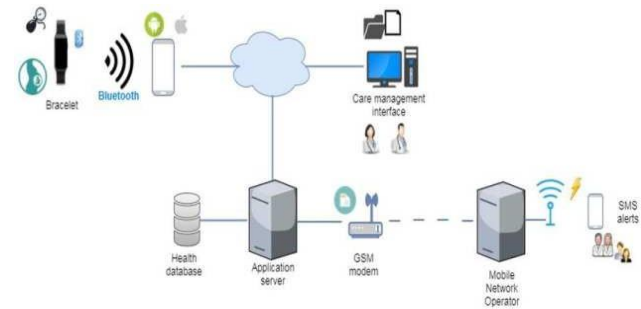

Fig. 1. *System architecture*

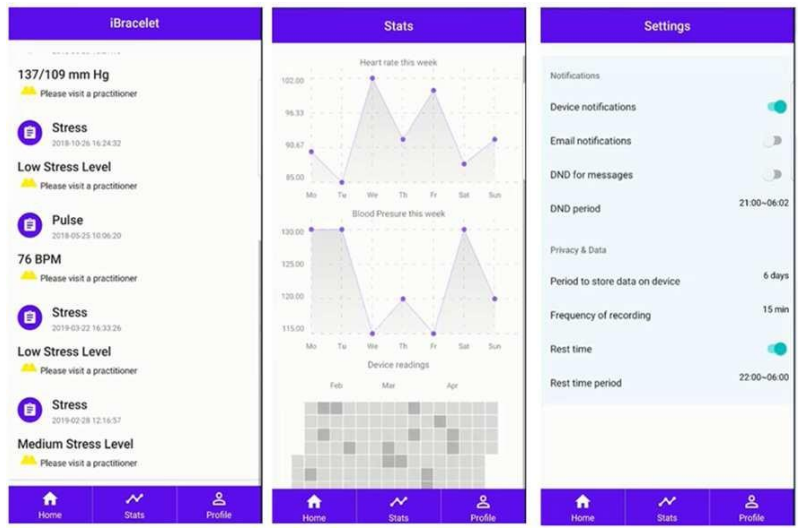

Fig. 2. *Mobile application interfaces*

The smart bracelet comprises a module with sensors, an electronic module and a wireless transmission module. The software application controls the system's operation and facilitates data communication, acquisition and processing.

The sensors detect the waveform of the blood pressure. The electronic module incorporates the conditioning and processing functionalities based on the input signal which is generated by the pressure sensor. In the next step, the output data is transmitted wirelessly to an associated smart phone. The dedicated algorithm processes the data and provides the relevant physiological parameters that represented numerically and graphically.

An alarm signal is triggered when the blood pressure value is over the normal level. The input data is provided by the system's smart bracelet which measures the blood pressure once at 15 minutes.

The measurement interval can be changed by the user of mobile phone application. The system's smart bracelet can be worn all day long, measuring continuously the blood pressure based on the artery of the hand, without the necessity of inflation.

### III. RESULTS

The i-bracelet software component uses the React-Native [25] and ElectronJS [26] technologies to make it available for Android and iOS mobile devices, while it can also be accessed via browser.

React-Native facilitates the creation of mobile phone applications using JavaScript and React as programming languages. The components are the same as for regular iOS and Android mobile applications, while they can also be written in Objective C, Java or Swift. The render is based on mobile user interface components. Recompiling is not needed, as the application reloads in real time.

ElectronJS is allows the programmer to build a website for a desktop application. In addition, it is a framework that is based on the JavaScript, HTML and CSS web technologies. ElectronJS uses Chromium and Node.js to build the application. This technology can be used by various operating systems, such as macOS, Windows and Linux. Updates are done automatically, notifications are received, reports are created if errors occur and debugging is available.

A dedicated REST API server for data storage has been developed for the described system. The software allows the user to manage the settings of the application. Their Bluetooth devices can be paired with the smart bracelet. In order to have access to the application's functionalities, the user has to register and authenticate. A feed with the latest in-app data is displayed when the application is opened. There is also a reporting module with interactive charts.

Fig. 2 illustrates the home screen with the blood pressure and pulse measurements, the stress level based on the number of steps counted by the smart phone, along with the screen where the statistics are illustrated. The last screen is for the profile where the application settings can be modified.

The statistics screen presents the evolution of the heart rate and blood pressure during the last week. The days colored in grey denote the case in which the values of the two measured parameters exceeded the threshold of 70 beats per minute (BPM), respectively 120/80 mmHg for the blood pressure.

The measurement results are displayed chronologically, as well as graphically. The evolution of the pulse and blood pressure are illustrated per week, as well as per month. For the per day chart, the last recorded value is displayed, as well as the values which exceed the threshold. If the last recorded value is above the threshold, then only that value appears. The charts show a detailed overview about the variation of the systolic and diastolic values, including the ones for the heart rate.

The settings which can be modified are the notifications for announcements on the device or to the user's email address. A do not disturb period can be set to determine the user's sleeping or relaxation hours. During this period, the incoming alerts can be disabled.

The period during which the measured records are stored on the phone can be changed, as well as the frequency of measurements. A rest time period can be set for studying the heart rate and the blood pressure values in different circumstances. During this time, the application analyzes the input data differently.

All these features have been developed for the life enhancement of the user. The graphical user interface is intuitive and easy to use. All the settings have been thought to offer the patients the chance to personalize the application's workflow according to their own lifestyles.

## IV. CONCLUSIONS

In this paper have been presented the main features of the mobile application which monitors the blood pressure measured values. The work was done in the context of the international i-bracelet Eurostars project.

The software component for the i-bracelet project was implemented using the React-Native and ElectronJS technologies, being available for Android and iOS mobile devices. The application is also accessible from a browser due to the facilities offered by the selected technologies.

The blood pressure and pulse values which are triggered by the system's smart bracelet are outlined in a written and visual manner. The values which exceed the normal threshold are mentioned and a suggestion for the user to visit a practitioner is provided every time the measured value exceeded its limit. Notifications are delivered when the critical state persists. By providing the sleep period, the monitoring of the blood pressure is enhanced in order to detect preeclampsia.

The presented solution is available as a mobile application, as well as an online website, and it offers the user the opportunity to analyze the evolution of the blood pressure and pulse rate. In this way, the user's lifestyle is respected and possible unpleasant situations can be avoided.

A follow up of the current work is the creation and enhancement of the artificial intelligence module for the automatic detection of preeclampsia.

## ACKNOWLEDGMENT

This work was funded by a grant of the Romanian National Authority for Scientific Research and Innovation, CCCDI - UEFISCDI, project number 59/2017, Eurostars Project E10871, i-bracelet— "Intelligent bracelet for blood pressure monitoring and detection of preeclampsia".